# Chaotic particle dynamics in free-electron lasers with coaxial wiggler

B. Farokhi and S. Mobarakabadi

Islamic Azad University of Arak, Arak, Iran

The motion of a relativistic test electron in a free-electron laser with an ideal coaxial-wiggler field and uniform axial-guide field are considered. We have investigated the group I, II, III orbits and finally have found that orbits become chaotic at sufficiently high beam density. We have changed beam radii, density of electron beam, intensity of coaxial magnetic field, and intensity of magnetic guide field as a parameter for finding chaotic area. An analytical estimate of the threshold value of the self-field parameter for the onset of chaos is obtained and found to be in good agreement with computer simulations. The threshold value of the wiggler amplitude for the onset of chaos is estimated analytically and confirmed by computer simulations for spatial case where self-field effects are negligibly small. Moreover, it is shown that the particle motion becomes chaotic on a time scale comparable with the beam transit time through a few wiggler periods.

### I. INTRODUCTION

Hamiltonian chaos has been an active area of research in physics and applied sciences. The classic work of kolmogovov, Arnold, and Moser (KAM) shows that the generic phase space of integrable classical Hamiltonian systems . subject to small perturbations, contains three types of orbits: stable periodic orbits, stable quasi periodic orbits, and chaotic orbits.

Earlier investigations of chaos in free – electron lasers have focused on chaotic behavior in particle orbits induced by sideband and radiation fields. Riyopoulos and Tang have analyzed side hand- induced chaos in the electron motion in the field configuration consisting of an ideal helical–wiggler field, the electromagnetic signal wave field, and the sideband wave field. Chen and Schmidt have shown that the electromagnetic signal wave can also cause chaotic electron motion in the combined helical-wiggler and axial guide field configuration. Then Billardon has observed evidence of chaotic behavior in the radiation field in a modulated storage rugfel. Michel-Lours *etal* are shown that the motion of an electron in a linearly polarized wiggler with an axial guide field is non integrable and chaotic. Chen and Davidson [5, 6] have found the electron dynamics in the self-magnetic field produced by the non-neutral electron beam in the field configuration consisting of a constant amplitude helical-wiggler magnetic field, and uniform axial magnetic field become chaotic.

In this paper, we analyze the motion of a relativistic electron in the field configuration consisting of a coaxial wiggler, and a uniform axial magnetic fields. It is shown that the motion is nonitegrable. Nonzero Lyapunov exponents are generated to demonstrate the nontegrablility and choaticity of the motion.

The organization of this paper is as follows. In see. II, theoretical formulation of the problem is analyzed. In see . III, the results of numerical computations and some conclusions are presented.

### II. THEORETICAL FORMULATION

The motion of one electron in a free electron laser (FEL) with coaxial wiggler $\mathbf{B}_w$ and a guide field $\mathbf{B}_0$ is considered. The self-field produced by the electron beam are neglected. The motion of the electron takes place in the coaxial wiggler. The total magnetic field inside a coaxial wiggler will be taken to be of the form

$$\mathbf{B} = B_r \hat{\mathbf{r}} + B_z \hat{\mathbf{z}}, \tag{1}$$

$$B_r = B_w F_r(r, z), \tag{2}$$

$$B_z = B_0 + B_w F_z(r, z), \tag{3}$$

where $B_0$ is a uniform static axial guide field, and $F_r$ and $F_z$ are known functions of cylindrical coordinates $r$ and $z$.

$$F_r = F_{r1} Sin(k_w z) + F_{r3} Sin(3k_w z), \tag{4}$$

$$F_z = F_{z1} Cos(k_w z) + F_{z3} Cos(3k_w z), \tag{5}$$

where

$$F_{rn} = G_n^{-1}[S_n I_1(nk_w r) + T_n K_1(nk_w r)], \quad (6)$$

$$F_{zn} = G_n^{-1}[S_n I_0(nk_w r) - T_n K_0(nk_w r)], \quad (7)$$

$$G_n \equiv I_0(nk_w R_{out})K_0(nk_w R_{in}) - I_0(nk_w R_{in})K_0(nk_w R_{out}), \quad (8)$$

$$S_n = \frac{2}{n\pi}Sin\left(\frac{n\pi}{2}\right) \times [K_0(nk_w R_{in}) + K_0(nk_w R_{out})], \quad (9)$$

$$T_n = \frac{2}{n\pi}Sin\left(\frac{n\pi}{2}\right) \times [I_0(nk_w R_{in}) + I_0(nk_w R_{out})], \quad (10)$$

and $n = 1,3$ ; $R_{in}$ and $R_{out}$ are the inner and outer radii of the coaxial waveguide, $k_w = 2\pi/l_w$ where $l_w$ is the wiggler (spatial) period, and $I_0$, $I_1$, $K_0$, and $K_1$ are modified Bessel functions. The corresponding Hamiltonian is.

As the Hamiltonian is not an explicit function of time, H is a constant of motion. Because H is independent of $\varphi$, it follows that $p_\varphi = Const$

For numerical calculation, dimensionless variables are introduced: $\hat{p}_i = p_i / mc$ , $\hat{z} = k_w z$ , $\hat{r} = k_w r$ , $\hat{\Omega}_c = \Omega_c / ck_w$ with $\Omega_c = eB_0 / m$ , $a_w = eB_w / mck_w$ , $\gamma = H / mc^2$ , $\tau = ck_w t$ .

The trajectory of an electron have plotted in the Ref. [7]. Chaos is, infact, confirmed by performing Poincare sections and calculating nonzero Lyapunov exponents. We tried to find a canonical transformation, for finding two other constants, but having failed in finding all constants f motion. Poincare maps have been plotted to demonstrate the non integrability of the motion. For this purpose, the following normalized equations of motion derived from Eq. (11), have been solved numerically

$$\dot{r} = \frac{\partial H}{\partial p_r} \quad , \dot{\varphi} = \frac{\partial H}{\partial p_\varphi} \quad , \dot{z} = \frac{\partial H}{\partial p_z}$$
$$\dot{p}_r = -\frac{\partial H}{\partial r} \quad , \dot{p}_\varphi = 0 \quad , \dot{p}_z = -\frac{\partial H}{\partial z} \quad (12)$$

The plane $(r, p_r)$ with $z = 0 \bmod 2\pi$ is chosen to be the Poincare surface of section. The numerical method used is a fourth order Runge-Kutta.

The existence of chaotic trajectories is confirmed by calculating nonzero Lyapunov exponents by two approaches. The first consists in considering two nearby trajectories with an initial tangential vector of norm $d_0$. The distance $d_n$ Between those trajectories is calculated numerically, and as soon as $d_n / d_0$ is greater than a quantity between 2 and 3, we renormalize $d_n$ to $d_0$. The Lyapunov exponent corresponding to the Poincare map is given by

$$\sigma = \lim_{\substack{t_{n\max} \to \infty \\ d_0 \to 0}} \frac{1}{t_{n\max}} \sum_1^{n\max} \log\left(\frac{d_n}{d_0}\right) \quad (13)$$

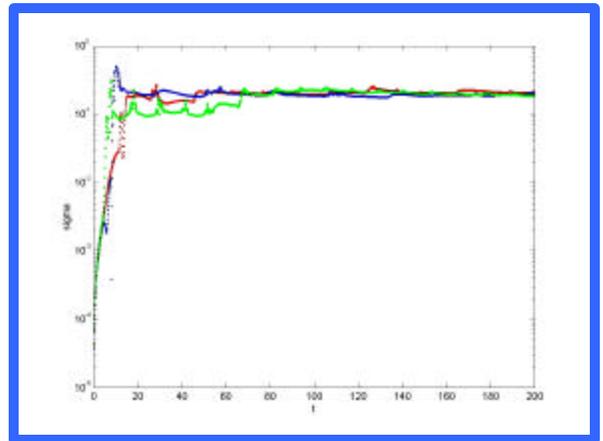

Figure: Lyapunov exponents corresponding to group III trajectory.